\documentstyle[twocolumn,revtex]{aps}

\def\ul#1#2{{\textstyle{#1 \over #2}}}

\def\o{{\it odd }}
\def\e{{\it even }}
\def\u{\uparrow}
\def\d{\downarrow}
\def\bk{{\bf k}}
\def\bp{{\bf p}}

\begin{document}
\draft

\widetext

\begin{title}
 Even and odd-frequency pairing correlations in 1-D t-J-h
model:\\ a comparative study\\

\end{title}

\author{A. V. Balatsky}
\begin{instit}
Center for Materials Science and Theoretical Division, T-11,\\ Los
Alamos National Laboratory, Los Alamos, NM 87545 and\\ Landau
Institute for Theoretical Physics, Moscow, Russia
\end{instit}
\author{J.\ Bon\v ca}
\begin{instit}
Theoretical Division, T-11, Center for Nonlinear Studies, \\ Los
Alamos National Laboratory, Los Alamos, NM 87545 and \\
J. Stefan Institute, University of Ljubljana, 61111 Ljubljana, Slovenia
\end{instit}

\receipt{\today}

\vfill

\begin{abstract}
An equal time version of odd-frequency pairing for a generalized $t-J$
model is introduced. It is shown that the composite operators
describing binding of Cooper pairs with magnetization fluctuations
naturally appear in this approach. The pairing correlations in both
BCS and odd-frequency channels are investigated exactly in 1D systems
with up to 16 sites. Our results indicate that at some range of
parameters odd-frequency correlations  become comparable, however
smaller than BCS pairing correlations. It is speculated that the spin
and density fluctuations in the frustrated model lead to the
enhancement of the \o gap susceptibilities.

\end{abstract}

\vfill


\newpage
{}~
\newpage

\narrowtext

The odd-frequency gap (\o gap hereafter) originally proposed for spin
triplet pairing by Berezinskii \cite{ber}, and recently extended on a
spin singlet pairing \cite{ba}, was the subject of a recent
investigation using the Monte Carlo method  on the Hubbard model
\cite{bsw}. The motivation for the current study was the observation
that \o pairing can be a possible pairing channel in strongly
correlated systems \cite{com1}.

As it was shown in \cite{G4}, the \o gap can be represented in another
form as an equal-time (i.e. even in frequency) composite four fermion
operator constructed out of equal time pairing field operator and a
magnetization operator (generally a particle-hole operator). The total
charge of this four fermion composite operator, i.e. the charge with
which this object transforms under electromagnetic $U(1)$, remains $q
= 2 e$ as it should. This representation offers  further insight
into the nature of condensate of \o pairing.

  To explore the possible existence of the odd-frequency gap SC
condensate in strongly correlated systems, we compared
quantitatively the \o and BCS pairing correlations in the recently
proposed 1D $t-J$ model in a staggered  $h$ field \cite{jb,jb1}.

One of the important advantages of studying 1D vs 2D models is that in
1D case we can numerically investigate much larger systems and thus
get information about pairing correlations at larger distances. For
example, in this study we report on a numerical calculation of both
BCS and \o pairing correlations in 1D system with up to 16 sites. This
should be compared with a typical Monte Carlo calculations in 2D of $6
\times 6$ sites. Large system size allows us to find pairing
correlations at much greater distances.  To make the physics of this
particular 1D model closer to real 2D physics we imposed the external
staggered magnetic filed. It has been shown in \cite{jb} that the
external staggered field $h$ in 1D simulates some 2D effects otherwise
not characteristic for ordinary 1D $t-J$ model. More precisely, it
induces a longer range antiferromagnetic ordering which leads to the
spin-string phenomenon \cite{bula}, strong mass renormalization
\cite{trug} and formation of bound hole pairs \cite{bonc}. The phase
diagram of the $t-J-h$ model shows three distinct regions \cite{jb1}.
When $J<J_c(h)$ there are no bound states, when $J_c(h)<J<J_s(h)$
system consists of bound hole pairs and when $J>J_c(h)$ phase
separation into a hole-rich and spin-rich phase occurs. Furthermore,
since $h\neq 0$ destroys the spin-rotation invariance of the system
and induces a gap in the spectrum of magnetic excitations, the model
belongs to the Luter Emery universality class \cite{emer}. The charge
exponent $K_\rho$, which determines the behavior of correlation
functions at large distances, was proven to be $K_\rho>1$ in a wide,
physically relevant regime of parameters, indicating on dominant SC
correlations \cite{jb1}.

In this paper we will report on a comparative study of \e (i.e. BCS)
and \o pairing correlations in spin singlet and triplet channels in 1D
$t-J-h$ model \cite{jb,jb1}, using Lanczos exact diagonalization
method.  First, we will derive an equal time version of \o gap and
show that: a) equal time order parameter describes the condensate of
composite operator, which is the product of equal time pair field and
magnetization operator; b) the equal time version of spin triplet \o
field operator is even $(P=+1)$ under parity transformation of
relative coordinates and is represented by a product of both spin
singlet and triplet equal time pairing field and appropriate
components of magnetization operator (see below). The spin singlet \o
operator can be written as a product of magnetization operator and
equal time spin triplet pairing fields and is odd ($P = -1$) under
spatial inversion. These features agree with general
symmetry requirements for \o singlet and triplet gaps \cite{ber,ba}.
Next, we compare pairing correlations in BCS and \o channel.  The BCS
correlations are always large in comparison with \o correlations in
the $t-J-h$ model at small hole-doping. However, in the extended
$t-t'-J-h$ model with next nearest neighbors hopping amplitude $t'$,
the \o pairing correlations increase and become comparable with the
BCS pairing correlations.

The main conclusion of this study is that in the pure $t-J-h$ model
the \o gap correlations are generically small. However, in the
extended $t-t'-J-h$ model these correlation are increased
substantially (by one order of magnitude). Moreover they become of the
same order as the largest BCS correlations for a particular sign of
$t'$. From our numerical calculations it also unambiguously follows
that the \o pairing correlation length is increasing rapidly with
increased hole doping, in contrast to suppression of the BCS
correlation length. This could indicate that \o pairing becomes the
dominant pairing channel close to phase separation region, where
particle and spin density fluctuations are strongly enhanced
\cite{ek}.

  We start with the hamiltonian of a generalized 1-D $t-J-h$ model
\cite{jb} in the following form

\begin{eqnarray}
H=&-&t\sum_{i\alpha}(\tilde c_{i,\alpha}^\dagger \tilde
c_{i+1,\alpha}+ H.c.) - t'\sum_{i\alpha}(\tilde c_{i,\alpha}^\dagger
\tilde c_{i+2,\alpha}+ H.c.)\nonumber\\
&+& J\sum_i (\vec S_i \cdot
\vec S_{i+1}- \ul14 n_i n_{i+1}) -
 h\sum_i (-1)^i S_i^z,
\label{eq1}
\end{eqnarray}
where $\tilde c_{i,\alpha} = c_{i,\alpha} (1-n_{i,-\alpha})$ is
projected fermionic operator onto the space of singly occupied sites,
$n_i$ is the corresponding fermion number operator, $\vec S_i$ is a
spin operator, $h$ is the strength of the external staggered field, and
the $t'$ term represents the next-neighbor hopping.  To simplify our
notation we omit in the rest of the paper the \~{ } sign above creation
and annihilation operators.

  First we consider a general two-particle gap function
\begin{equation}
\Delta_{\alpha\beta}(\bk,2\tau) =\langle T_\tau
c_{\alpha,\bk}(\tau) c_{\beta,-\bk}(-\tau) \pm (\tau \to -\tau)\rangle,
\label{eq2}
\end{equation}
where $T_\tau$ is Matsubara time ordering operator and we assume
pairing in the center of mass momentum zero. We constructed the gap
function in (\ref{eq2}) to be explicitly even ($+$ sign) or odd ($-$
sign) in the imaginary time $\tau$. As recently suggested
\cite{ba,ber}, the only requirement for pairing in the singlet channel
is that the gap function $\Delta_{\alpha\beta}(k,\tau)$ is even under
simultaneous space and imaginary-time reversal, i.e. $\bk\to -\bk,
\tau \to -\tau$.  Consequently, in the case of the singlet pairing,
the gap function can be either even in both space and imaginary-time,
leading to the conventional BCS gap, or can be odd in both space and
imaginary-time, leading to the \o gap.  Furthermore, in the case of
triplet pairing the gap function must be odd under simultaneous space
and imaginary-time reversal again leading to two solutions: a
conventional BCS where gap function is even in $\tau$ and odd in $k$
or \o gap, which is odd in $\tau$ and even in $k$.

  We are mainly interested in the lowest order contributions to the
gap in the relative pair-field time $2\tau$. Then, assuming
analyticity of $\Delta_{\alpha\beta}$ in small $\tau$, we have
\begin{eqnarray}
\Delta_{\alpha\beta}^{\it even}&=& \Delta_{\alpha\beta}^{(0)} +
{\cal O}(\tau ^2)\label{eq3a}\\
\Delta_{\alpha\beta}^{\it odd}&=& 2\tau \Delta_{\alpha\beta}^{(1)} +
{\cal O}(\tau ^3),
\label{eq3b}
\end{eqnarray}
where the equal-time contribution in Eq.(\ref{eq3a}) corresponds to
the {\it even} gap, which is the conventional BCS gap, whereas the
higher order corrections in $\tau$ generate further multi-particle
dressing. By definition there is no zeroth order term in $\tau$ (i.e.
equal-time pairing) term in the case of \o gap Eq.(\ref{eq3b}).  To
calculate the lowest equal time pairing field associated with
odd-pairing we take the time derivative of the gap function
\begin{eqnarray}
{\partial\Delta_{\alpha\beta}(\bk,2\tau)\over\partial
\tau}\vert_{\tau=0}= \Delta^{(1)}_{\alpha\beta}(\bk,0)=\nonumber\\
\langle
 \dot c_{\alpha,\bk} c_{\beta,-\bk}- c_{\alpha,\bk} \dot c_{\beta,-\bk}\rangle,
\label{eq4}
\end{eqnarray}
where $c_{\gamma,\bp}, \dot c_{\gamma,\bp}$ are calculated at $\tau = 0$.

The odd-$\tau$ pair-field operator can be written in the following
form
\begin{equation}
\Delta \propto \sum_\bk \Delta^{(1)}_{\alpha\beta}(\bk)
(\dot c_{\alpha,\bk}c_{\beta,-\bk}-
c_{\alpha,\bk}\dot c_{\beta,-\bk}).
\label{eq5}
\end{equation}
We are still free to choose  $\Delta^{(1)}_{\alpha\beta}(\bk)$ to
be an even  (odd) ($P = \pm 1$) under space transformation $\bk \to
-\bk$, thus obtaining a triplet (singlet) \o gap pair-field operator.
To rewrite Eq.~(\ref{eq5}) in a more explicit form we proceed by
calculating equations of motion for a Hamiltonian of $t-t'-J-h$ model
in the $r$-space
\begin{eqnarray}
\dot{c}_{i,\alpha} = -\vec M_i \vec\sigma_{\alpha\beta}c_{i,\beta} -h
(-1)^i\sigma^z_{\alpha\alpha} c_{i,\alpha}\nonumber\\
+t(c_{i-1,\alpha} +
c_{i+1,\alpha}) + t'(c_{i-2,\alpha} + c_{i+2,\alpha}),
\label{eq6}
\end{eqnarray}
with $\vec M_i= J(\vec S_{i-1} + \vec S_{i+1})$ and $\vec
\sigma_{\alpha\beta}$ are Pauli matrices.  Assuming that the gap
function $\Delta^{(1)}_{\u\d}(k)$ is $\sin k$ in the case of \o
singlet pairing and $\cos k$ for the $S_z = 0$ triplet and that
$\Delta^{(1)}_{\u\u}(k) = \cos k$ for $S_z = 1$ triplet, we obtain
after a Fourier transformation from Eq.~(\ref{eq6}) and
Eq.~(\ref{eq4}) the following compact form of \o singlet and triplet
pairing operators:
\begin{eqnarray}
\Delta^{odd}_{singlet}(r_i) &\propto & \left ( \vec S_{i-1}+\vec S_{i+2}
\right )
\left( \sigma^y\vec \sigma\right)_{\alpha\beta}c_{i,\alpha}c_{i+1,\beta}
\label{eq7a}\\
\Delta^{odd}_{triplet,S_z=0}(r_i) &\propto &
\left (\vec S_{i-1}\left( \sigma^y\vec \sigma\right)_{\alpha\beta}
-\vec S_{i+2}\left( \sigma^y\vec \sigma\right)_{\beta\alpha}\right)\nonumber\\
\times c_{i,\alpha}c_{i+1,\beta}
\label{eq7b}\\
\Delta^{odd}_{triplet,S_z=\pm 1}(r_i) &\propto &
\left ( \vec S_{i-1}\left(\vec \sigma \pm\sigma^z\vec \sigma\right)_
{\alpha\beta}
-\vec S_{i+2}\left(\vec\sigma\pm\sigma^z\vec \sigma\right)_
{\beta\alpha}\right)\nonumber\\ \times c_{i,\alpha}c_{i+1,\beta}
\label{eq7c}
\end{eqnarray}

In deriving Eqs.~(\ref{eq7a},\ref{eq7b},\ref{eq7c}) the terms
proportional to $t,t'$ and $h$ drop out, since they contribute only
to the \e pairing operators.

It is instructive to investigate the composite structure of \o pairing
operator from Eqs.~(\ref{eq7a},\ref{eq7b},\ref{eq7c}) more closely.
Consider $\Delta^{odd}_{singlet}$, which is a product of magnetization
operator $( \vec S_{i-1}+\vec S_{i+2})$ with the BCS spin triplet
pairing field $\vec{\Delta}^{even}_{triplet} \propto \left(
\sigma^y\vec
\sigma\right)_{\alpha\beta}c_{i,\alpha}c_{i+1,\beta}$. Since product
of these two operators form a scalar in spin space, it immediately
follows that $\Delta^{odd}_{singlet}$ is indeed a singlet. The
spatial parity of composite operator is the product of parity of
magnetization operator $P_{magn} = +1$: at $i-1 \leftrightarrow i+2
$ we have $(\vec S_{i-1}+\vec S_{i+2}) \leftrightarrow (\vec S_{i-1}+\vec
S_{i+2}) $ and  the parity of $\vec{\Delta}^{even}_{triplet}$, which
 is $P_{BCS}= -1$. Thus it follows that the composite \o operator
from Eq.~(\ref{eq7a}) represents the odd parity $P = -1$ spin singlet
pairing filed. The analogous consideration can be done for \o triplet
operator as well.

{}From the structure of composite operators
Eqs.~(\ref{eq7a},\ref{eq7b},\ref{eq7c}) it is clear that the \o
pairing correlations will be suppressed due to phase space
restrictions. However, there is an important caveat: close to the
phase separation region in the $t-J$ model, spin and density
fluctuations are strongly enhanced at intermediate length scale
\cite{ek}. A snapshot of the system will show metallic regions with a
large density of holes and antiferromagnetic regions with less holes.
These fluctuations are soft and their spectral density $A(\omega)$ has
a strong peak at small frequencies. Under these circumstances the soft
spin boson attached to the Cooper pair in the composite operator is
readily available in the system and the phase space restrictions
become less important.

{}From this argument we should expect that the \o pairing
susceptibilities will increase with frustration in $t-J$ model. We
indeed find some enhancement of \o pairing correlations in a
frustrated $t-t'-J$ model, as we will show below.

  We will now turn to our numerical results. We solved the
$t-J-h$ model on a chain with periodic boundary conditions using the
standard Lanczos technique. We restricted our calculations to the
systems with $N=16~N_h=2$, $N=14~ N_h=4$ and $N=12~N_h=6$, where $N$
represents the number of sites and $N_h$ the number of holes in the
system.  In all cases investigated the ground state has the quantum
numbers $S_z=0, k=0$. Note, that due to the staggered external field
the total spin is no longer a good quantum number.

We searched for the most favorable pairing channel using the equal-time
pair-field susceptibility in the following form
\begin{equation}
P(r) = {\sum_{i=1}^N \langle\Delta^\dagger(r_i+r)\Delta(r_i)\rangle \over
\sum_{i=1}^N \langle\Delta^\dagger(r_{i})\Delta(r_i)\rangle},
\label{eq8}
\end{equation}
where we took for $\Delta(r_i)$ the \o gap,
Eqs.~(\ref{eq7a},\ref{eq7b},\ref{eq7c}) or BCS gap pairing fields. We
used a standard definition for the BCS pairing fields:
$\Delta^{BCS}(r_i)\propto c_{\u,i}c_{\d,i+1}\pm c_{\d,i}c_{\u,i+1}$
where - (+) represents singlet (triplet) $S_z=0$ pairing field and
$\Delta^{BCS}_{trilet,S_z=1}(r_i)=c_{\u,i}c_{\u,i+1}$ is a triplet
$S_z=1$ pairing-field.

  In Fig.~(1) we present density-density correlation functions
$g(r)=\sum_i\langle n_{h,i} n_{h,i+r}\rangle$, where $n_{h,i}=1-n_i$,
in the system of $N=14$ sites and $N_h=4$ holes at $h/t=1.0$ and two
different values of $J$. At $J/t = 0.5$ (open circles) $g(r)$ exhibits
a peak at the largest possible distance taking into account periodic
boundary conditions $r_m=N/2$ indicating that no bound hole-pairs are
in the system. However, at a greater value of $J/t=2.25$ there are two
peaks in $g(r)$ at $r=1$ and $r=r_m$ which are consistent with the
picture of two separate bound hole-pairs.

    In Fig.~(2a,b) BCS (open symbols) and \o gap (filled symbols)
pairing susceptibilities $P$ are shown as functions of distance $r$
between pairs  for the same choice of parameters as in Fig.~(1).
At $J/t=0.5$, Fig.~(2a), where there are no bound pairs in the system,
pairing susceptibilities rapidly decay with the distance. This rapid
decay is also in agreement with the value of the charge exponent
$K_\rho$ being less than unity \cite{jb1}.  In Fig.~(2b) we
present pairing susceptibilities at $J/t=2.25$.  We observe strong
enhancement of the BCS pairing susceptibilities at larger distances,
however, there is no major enhancement in the \o gap channel. The
enhancement of the pairing susceptibilities is consistent with the
formation of bound hole-pairs, enhancement of the charge exponent
$K_\rho>1$ and the proximity of the phase separation
\cite{jb1}.

  In our search for a possible enhancement of \o gap pairing
susceptibilities we also included  a second-neighbour hopping term
$t'/t$. The effect of the next-neighbor hopping is strongly
asymmetric.  Negative $t'/t<0$ leads to: a substantial spin-charge
coupling (nearly absent in the 1D $t-J$ model), formation of
ferromagnetic polarons and finally phase separation \cite{sega}.
Since finite $h>0$ enhances antiferromagnetic type of spin ordering,
we argue that $t'/t<0$ induces intability in the $t-t'-J-h$ model. On
the contrary, the effect of $t'/t>0$ on the pairing correlations is
much less pronounced. In Fig.~(2) we present the comparison of largest
pairing susceptibilities in the BCS and \o gap channel at $J/t = 1.5,
h/t = 1.5$ and two different values of $t'/t=0, -0.3$. Clearly there
is an opposite,  weak effect of $t'$ on pairing
susceptibilities. On the one hand, the BCS pairing correlations
diminish when $t'$ is switched on; on the other hand, we observe small
but consistent increase of \o gap susceptibility at large distances.

   Finally, we investigated the effect of doping on pairing in \o and
BCS channel. To present results in a more compact form we define two
quantities: $P_\Sigma$, representing a sum over pairing off-diagonal
susceptibilities $P_\Sigma = \sum_r \vert P(r)\vert/N$, where we
choose $r\geq 1$ to avoid overlap of nonlocal operators with
themselves.  Consider also the first moment of $P(r)$ which is defined
as $\xi = \sum_r r\vert P(r)\vert/N\sum_r\vert P(r)\vert $ and gives
the correlation length of superconducting fluctuations. In Table we
present results for $P_\Sigma$ and $\xi$ for three different systems
with $N=16~N_h=2, N=14~N_h=4$ and $N=12~N_h=6$, at $J/t = 1.0, h/t =
1.0$ and $t'=0$.  The presented $P_\Sigma$ correspond to singlet
pairing in both \o and BCS channel, since at given parameters singlet
pairing susceptibilities yield maximum values of $P_\Sigma$. While
$P_\Sigma$ is increasing with hole-doping $\eta=N_h/N$ in both
channels, the correlation length $\xi$ displays different behavior. In
the BCS channel $\xi$ first decreases while hole doping decreases from
$\eta=2/16=0.125$ to $4/14\simeq 0.286$ and then slightly increases at
$\eta = 6/12=0.5$.  In the \o channel $\xi$ exhibits more than a
twofold monotonic increase with doping.

 As was argued in the second paragraph after
 Eqs.~(\ref{eq7a},\ref{eq7b},\ref{eq7c}), the substantial increase in
$\xi$ with doping in the \o channel of the $t-J-h$ model is the
consequence of the strongly enhanced spin fluctuations at low
frequencies. This result might indicate that frustration due to
hole-doping leads to a long-range order in the \o pairing channel.

\underline{In conclusion}, we derived  composite operators describing
the condensate of superconductor with \o gap for a generalized $t-J$
model. These composite operators always involve a bound state of Cooper
pair and a proper combination of magnetization operators.  Due to the
presence of an extra spin 1 boson in composite operator the parity in
\o gap is opposite to that in BCS case, e.g. odd parity singlet in \o
case (see Eq.~(\ref{eq7a})) vs.  even parity singlet in BCS. The BCS
and \o pairing correlations were investigated numerically for the 1D
$t-t'-J$ model in external staggered field. The enhancement of the BCS
correlations coincides with the formation of bound hole pairs and with
charge exponent being $K_\rho >1$ in the $t-J-h$ model.  It is shown
that at small hole doping and when $t'=0$ the strongest \o pairing
correlations are always smaller then strongest BCS correlations.
However, despite the phase space restrictions, the \o correlations
increase substantially with hole doping and when $t'<0$ in a {\it
frustrated} $t-t'-J-h$ model. We argue that doping and frustration
create additional soft spin fluctuations what makes \o correlations
more pronounced, whereas BCS correlations are suppressed by doping.

\underline{Acknowledgments} We are grateful to authors of Ref.\cite{G4} for
useful discussions. One of the authors (J.B.) benefited from
discussions with P.~Prelov\v sek and is in particular grateful to
I.~Sega who suggested the compact form of \o gap pairing correlations,
presented in this paper. This work was supported by J.  R.  Oppenheimer
fellowship (A.B) and by Department of Energy. Part of this work was
done at Aspen Center for Physics, whose support is acknowledged.

\eject

\narrowtext
\begin{table}
\setdec 0.0000
\caption{Values for $P_\Sigma$ and correlation length $\xi$ for
singlet BCS and singlet \o pairing correlations are given for three
different values of doping $\eta = 0.125, 0.286, 0.5$ in $t-J-h$ model
at $J/t = 1.0, h/t = 1.0$. The pairing correlations in triplet channel
are smaller.  }
\begin{tabular}{lcccc}
       &\multicolumn{2}{c} {BCS} &\multicolumn{2}{c} {ODD}\     \\
$\eta$ & $P_\Sigma$  & $\xi$     & $P_\Sigma$  & $\xi$         \\ \tableline
0.125  &\dec 0.0369 &\dec 0.4487 &\dec 0.0318 &\dec 0.1840 \\
0.286  &\dec 0.0840 &\dec 0.2615 &\dec 0.0537 &\dec 0.2102 \\
0.500  &\dec 0.1924 &\dec 0.2864 &\dec 0.0838 &\dec 0.4100 \\
\end{tabular}
\label{table1}
\end{table}

\figure{Hole-density correlation function $g(r)$ vs. distance $r$
in the $t-J-h$ model, for $N=14, N_h=4$ and two different values of
$J/t$.}

\figure{Pairing susceptibilities $P$ vs. distance $r$ in the $t-J-h$
model. Parameters of the model are the same as in the Fig.~(1). Note,
that the legend in Fig.~(2a) is also valid for Fig.~(2b). Pairing
susceptibilities are presented only for $r\geq 2$ since at smaller
distances pairs overlap.}

\figure{The largest pairing susceptibilities in the BCS and \o channel
in the $t-t'-J-h$ model for $N=14, N_h=4$. Circles present \o singlet
pairing, squares present BCS triplet $S_z=0$ pairing.}

\end{document}